\documentclass[conference,compsoc,onecolumn]{IEEEtran}
\ifCLASSOPTIONcompsoc
  \usepackage[nocompress]{cite}
\else
  \usepackage{cite}
\fi

\ifCLASSINFOpdf
\else
\fi
\usepackage{amsmath}

\usepackage{array}

\ifCLASSOPTIONcompsoc
\usepackage[caption=false,font=footnotesize,labelfont=sf,textfont=sf]{subfig}
\else
\usepackage[caption=false,font=footnotesize]{subfig}
\fi
\usepackage{url}

\usepackage{graphicx}

\usepackage{algorithm,algpseudocode}
\algrenewcommand\algorithmicrequire{\textbf{Input:}}
\algrenewcommand\algorithmicensure{\textbf{Output:}}
\usepackage{amsmath,amsfonts,amssymb,amsthm}
\usepackage{xcolor}
\definecolor{ocre}{RGB}{0,173,239}

\usepackage{tikz}
\usepackage{pgfplots}
\pgfplotsset{compat=1.18}

\usetikzlibrary{arrows,matrix,positioning,fit}
\usetikzlibrary{patterns,decorations.pathreplacing}
\tikzset{%
  highlight/.style={rectangle,rounded corners,fill=ocre!50,draw=ocre!80,
     fill opacity=0.2,inner sep=0pt,text opacity=1}
}

\usepackage{graphicx} 
\usepackage{float}
\usepackage[all]{xy}
\usepackage{tikz-cd}
\usepackage[shortlabels]{enumitem}
\usepackage{dsfont}
\usepackage{tikz}
\usepackage{pbox}
\usetikzlibrary{backgrounds}
\pgfdeclarelayer{background2}
\pgfdeclarelayer{background}
\pgfdeclarelayer{foreground}
\pgfdeclarelayer{foreground2}
\pgfsetlayers{background2,background,main,foreground,foreground2}

\makeatletter
\renewcommand*\env@matrix[1][*\c@MaxMatrixCols c]{%
  \hskip -\arraycolsep
  \let\@ifnextchar\new@ifnextchar
  \array{#1}}
\makeatother

\usepackage[edges]{forest}
\usepackage{array}
\newcolumntype{C}[1]{>{\centering\arraybackslash}p{#1}}
\newcolumntype{L}[1]{>{\raggedright\arraybackslash}p{#1}}
\usetikzlibrary{arrows.meta,shadows}
\usetikzlibrary{trees,positioning,shapes,shadows,arrows}

\usepackage[hypertexnames=false]{hyperref}
\hypersetup{pdfborder=0 0 0}

\algnewcommand{\LineComment}[1]{\State \(\triangleright\) #1}

\usepackage{tablefootnote} 

\usepackage{lastpage} 
\usepackage{fancyhdr} 

\newlength{\gpgpuElemSep}
\newlength{\gpgpuElemSize}
\usepackage{listings}
\usepackage{caption}
\makeatletter
\let\old@lstKV@SwitchCases\lstKV@SwitchCases
\def\lstKV@SwitchCases#1#2#3{}
\makeatother
\usepackage{lstlinebgrd}
\makeatletter
\let\lstKV@SwitchCases\old@lstKV@SwitchCases

\lst@Key{numbers}{none}{%
    \def\lst@PlaceNumber{\lst@linebgrd}%
    \lstKV@SwitchCases{#1}%
    {none:\\%
     left:\def\lst@PlaceNumber{\llap{\normalfont
                \lst@numberstyle{\thelstnumber}\kern\lst@numbersep}\lst@linebgrd}\\%
     right:\def\lst@PlaceNumber{\rlap{\normalfont
                \kern\linewidth \kern\lst@numbersep
                \lst@numberstyle{\thelstnumber}}\lst@linebgrd}%
    }{\PackageError{Listings}{Numbers #1 unknown}\@ehc}}
\makeatother

\def\FunctionD(#1){2*(#1)-1}%
\def\FunctionM(#1,#2){((3*(#1)^2)*10^9)/(#2)}
\newcommand{\PreserveBackslash}[1]{\let\temp=\\#1\let\\=\temp}
\newcommand\hlightcolor[1]{\tikz[overlay, remember picture,baseline=-\the\dimexpr\fontdimen22\textfont2\relax]\node[rectangle,fill=magenta!50,rounded corners,fill opacity=0.2,draw=magenta!80,text opacity=1] {$#1$};} 
\usepackage{soul} 
\setstcolor{blue}
\usepackage{accents}

\definecolor{myblue4} {RGB}{66,129,196}
\tikzstyle{myline} = [very thick, draw=myblue4, fill=myblue4, shape=rectangle, inner sep=10pt, inner ysep=20pt]
\tikzstyle{mybox} = [very thick, draw=myblue4, fill=white, shape=rectangle, inner sep=10pt, inner ysep=20pt]

\tikzstyle{mybox1} = [very thick, draw=white, fill=white, shape=rectangle, inner sep=10pt, inner ysep=20pt]

\usepackage{adjustbox}
\usepackage{fourier-orns}
\usepackage{multirow}
\usepackage{tikz}
\usepackage{pgfplotstable}
\usepackage{ifthen}

\DeclareCaptionFormat{algor}{%
\hrulefill\par\offinterlineskip\vskip1pt%
\textbf{#1#2}#3\offinterlineskip\hrulefill}
\DeclareCaptionStyle{algori}{singlelinecheck=off,format=algor,labelsep=space}
\usepackage{bm}
\usepackage{setspace}

\usepackage{multicol}
\usepackage{pdflscape}

\usepackage{longtable}
\usepackage{stackengine}

\usepackage{subcaption}
\selectcolormodel{gray}
\pgfplotscreateplotcyclelist{my_black_white}{%
solid, every mark/.append style={solid, fill=gray}, mark=*\\%
solid, every mark/.append style={solid, fill=gray}, mark=square*\\%
solid, every mark/.append style={solid, fill=gray}, mark=star\\%
solid, every mark/.append style={solid, fill=gray}, mark=diamond*\\%
solid, every mark/.append style={solid, fill=gray}, mark=otimes\\%
solid, every mark/.append style={solid, fill=gray}, mark=Mercedes star\\%
solid, every mark/.append style={solid, fill=gray}, mark=triangle*\\%
solid, every mark/.append style={solid, fill=gray}, mark=oplus\\
densely dotted, every mark/.append style={solid, fill=gray}, mark=*\\%
densely dotted, every mark/.append style={solid, fill=gray}, mark=square*\\%
densely dotted, every mark/.append style={solid, fill=gray}, mark=star\\%
densely dotted, every mark/.append style={solid, fill=gray}, mark=diamond*\\%
densely dotted, every mark/.append style={solid, fill=gray}, mark=otimes\\%
densely dotted, every mark/.append style={solid, fill=gray}, mark=Mercedes star\\%
densely dotted, every mark/.append style={solid, fill=gray}, mark=triangle*\\%
densely dotted, every mark/.append style={solid, fill=gray}, mark=oplus\\
dashed, every mark/.append style={solid, fill=gray}, mark=*\\%
dashed, every mark/.append style={solid, fill=gray}, mark=square*\\%
dashed, every mark/.append style={solid, fill=gray}, mark=star\\%
dashed, every mark/.append style={solid, fill=gray}, mark=diamond*\\%
dashed, every mark/.append style={solid, fill=gray}, mark=otimes\\%
dashed, every mark/.append style={solid, fill=gray}, mark=Mercedes star\\%
dashed, every mark/.append style={solid, fill=gray}, mark=triangle*\\%
dashed, every mark/.append style={solid, fill=gray}, mark=oplus\\
dashdotted, every mark/.append style={solid, fill=gray}, mark=o\\%
dashdotted, every mark/.append style={solid, fill=white}, mark=square*\\%
}
\begin{document}
\pagenumbering{arabic}
\pagestyle{plain}

\title{Multi-GPU Tridiagonal Partition Algorithm}

\author{
\IEEEauthorblockN{Milena Veneva\IEEEauthorrefmark{1}, \emph{milena.p.veneva@gmail.com}}
\IEEEauthorblockA{\IEEEauthorrefmark{1}RIKEN Center for Computational Science, R-CCS, 7-1-26 Minatojima-minami-machi, Chuo-ku, \\Kobe, Hyogo 650-0047, Japan}
} 

\maketitle
\thispagestyle{plain}

\section*{Abstract}
This paper presents a multi-GPU version and a multi-GPU recursive version of the 
parallel partition method for systems of linear algebraic equations with 
tridiagonal coefficient matrices and their implementations in CUDA. Computational 
experiments on one and two GPU cards, for different system sizes and with 
different number of recursive steps are conducted. The results are analyzed and 
discussed. 
\section{Introduction}
The parallel partition algorithm for solving systems of linear algebraic 
equations~(SLAEs) with tridiagonal coefficient matrices proposed 
in~\cite{Austin_2004} represents a conceptually simple yet highly efficient 
numerical method. It operates by decomposing the original matrix into smaller 
sub-matrices, thereby enabling the concurrent solution of smaller SLAEs. The 
algorithm was originally designed for execution on systems with a large number 
of processors and accordingly, the computational experiments reported 
in~\cite{Austin_2004} use up to 250 processors.  

The algorithm operates in three stages. In Stage 1, we partition the initial 
SLAE into sub-systems and compute two interface equations for each. In Stage 2, 
we assemble these interface equations into a tridiagonal interface system, 
which maintains diagonal dominance if the original system possesses it. We solve 
the interface system and thus, find a subset of the initial unknowns. Finally, 
in Stage 3 these values are substituted and the remaining now independent 
sub-systems are solved simultaneously in parallel.

We have implemented this algorithm in CUDA~\cite{cuda_progr_guide}. Previously 
we have built heuristics for finding the optimum number of CUDA streams for a 
particular SLAE size in~\cite{Veneva_2025b}, the optimum number of recursive 
steps and the optimum sub-system size  for a particular SLAE size 
in~\cite{Veneva_2026}. The tridiagonal partition algorithm contains enough 
parallelism to justify the usage of multiple GPUs to improve performance, 
and/or to help handling SLAEs which are too big to be solved on a single GPU due 
to global memory limitations. A multi-GPU CUDA algorithm and a multi-GPU 
recursive CUDA algorithm of the partition method are built and implemented in 
CUDA.
\section{Multi-GPU algorithm and CUDA implementation}

To run the code on multiple GPUs, the needed device memory should be allocated 
across the available devices, and the SLAE should be partitioned into as many 
chunks as the number of the these available GPUs. All this, together with the
copy-compute overlap approach on which the partition implementation relies, 
requires careful calculation of the sizes of the data chunks which are going to 
be send to each device, the sizes of the data chunks that are going to be 
processed by a single CUDA stream on each of the devices, and the number of CUDA 
blocks that are going to be used on each stream. 
In our case, as each SLAE partition can be run independently on a GPU, the GPUs 
do not need to communicate between each other during the computation.
In the case of a computer system with $p$ equivalent GPU devices, the 
initial SLAE with size $N$ is split by integer division into $p$ 
sub-systems, each with size (approximately) $N/p$. This is 
aimed to keep all GPUs equally busy~\cite{Hart_2024}. The algorithm is 
going to work on a computer system which is equipped with different GPU cards as 
well, but then we cannot guarantee lack of idle times on some of the GPUs. 
Currently, the multi-GPU parallel 
partition method implementation works only on a single node. If we want to make 
it work with GPUs which reside on multiple nodes across the computer system, we 
would need to use MPI so as to handle the GPU~affinity (see, for 
instance,~\cite{Kraus_2022}). Since all operations in a non-default stream are 
non-blocking with respect to the host, the CUDA API function 
cudaStreamSynchronize() is called on each GPU to block the host thread until all 
previously issued operations in the respective stream on each GPU have 
completed.

The single-GPU implementation starts with sequential host-to-device~(H2D) 
memory transfer of the three arrays that hold the three diagonals of the SLAE. 
Then, in a for-loop we apply copy-compute overlap of the H2D memory transfer of 
the right-hand side of the SLAE, but in chunks, the kernel which is responsible 
for Stage~1, and the device-to-host~(D2H) memory transfers of the result of 
Stage~1, again in chunks. On the 
other hand, in the multi-GPU implementation we move the sequential H2D 
memory transfer of the three arrays that hold the three diagonals of the SLAE 
within the for-loop, and thus all the four H2D memory transfers are done in 
chunks. These two approaches are giving the best computational times for the 
partition algorithm, which is memory-bound, because: (1) in the single-GPU case 
we must minimize the per-chunk data size to prevent the kernel execution engine 
from too big idle times, and (2) in the multi-GPU case we need to utilize the 
multiple separate hardware pipelines effectively multiplying our system's 
data-ingestion speed. 
The pageable memory leads to serialization of the memory transfer in 
comparison with the pinned one. Hence, we use pinned memory.

\subsection{Computational experiment. }

The computational experiment is summarized as of one and two NVIDIA GPU RTX 
A5000~\cite{a5000a}, for SLAE sizes $10^{i}$, $i = 5, 6, 7, 8$, 
the number of CUDA streams was determined according to the optimum stream 
heuristic~\cite{Veneva_2025b}, 256 CUDA threads per block, FP64 precision, 
out-of-place solution (meaning that we do not overwrite the right-hand side of 
the initial SLAE), 0, 1, 2, and 3 recursions, the corresponding optimum 
sub-system sizes for these number of recursive steps -- $m$ (for the initial 
SLAE), $m_{1}$ (for the interface SLAE obtained after Stage 1 has been applied 
once), $m_{2}$ (for the interface SLAE obtained after Stage 1 has been applied 
twice), $m_{3}$ (for the interface SLAE obtained after Stage 1 has been applied 
three times) were determined according to the optimum sub-system size 
heuristic~\cite{Veneva_2026}, and are listed in Table~\ref{tab:mgpus_sub-system_sizes}.
\begin{table}[!htbp]
\centering
\caption{Sub-system sizes $m$, $m_{1}$, $m_{2}$, $m_{3}$ used in the 
computational experiments on one and two 
NVIDIA GPU RTX A5000, for 0, 1, 2, and 3 recursions, SLAE sizes $N = 10^{5}$, 
$10^{6}$, $10^{7}$, $10^{8}$.}
\label{tab:mgpus_sub-system_sizes}
\resizebox{1\textwidth}{!}{%
\begin{tabular}{|l|l|l|l|l|l|}
\hline
recursions & sub-system sizes & \multicolumn{1}{|c|}{$N = 10^{5}$} & \multicolumn{1}{|c|}{$N = 10^{6}$} & \multicolumn{1}{|c|}{$N = 10^{7}$} & \multicolumn{1}{|c|}{$N = 10^{8}$} \\\hline
0 & $m$ & 32 & 64 & 64 & 64 \\\hline
1 & $m$, $m_{1}$ & 32, 5 & 64, 20 & 64, 40 & 64, 32 \\\hline
2 & $m$, $m_{1}$, $m_{2}$ & 32, 5, 5 & 64, 5, 5 & 64, 10, 20 & 64, 10, 32 \\\hline
3 & $m$, $m_{1}$, $m_{2}$, $m_{3}$ & 32, 10, 4, 4 & 64, 10, 8, 8 & 64, 10, 20, 4 & 64, 5, 40, 20 
\\\hline
\end{tabular}}
\end{table}

Table~\ref{tab:recursive_1} summarizes the computational times (in [ms]) on one 
GPU, and additionally provides the times for the serial Thomas method 
(penultimate column) and the cuSPARSE tridiagonal solver~\cite{CUSPARSE} 
(last column). Table~\ref{tab:recursive_MGPUs_pinned} sums up the computational 
times (in [ms]) on two GPUs. In both the tables the best time is given in bold font. 
\begin{table}[!htbp]
\centering
\caption{Computational times (in [ms]) for the partition recursive methods on 
one NVIDIA GPU RTX A5000. The best time is given in bold font.}
\label{tab:recursive_1}
\resizebox{1\textwidth}{!}{%
\begin{tabular}{|c|c|c|c|c|c|c|c|}
\hline
\multicolumn{1}{|c|}{SLAE} & 
\multicolumn{1}{|c|}{number of} & 
\multicolumn{4}{|c|}{computational time for the partition method with}
& Thomas & cuSPARSE\\ \cline{3-6}
\multicolumn{1}{|c|}{size $N$} & \multicolumn{1}{|c|}{CUDA streams} & 
\multicolumn{1}{|c|}{0 recursions} & \multicolumn{1}{|c|}{1 recursion} & 
\multicolumn{1}{|c|}{2 recursions} & \multicolumn{1}{|c|}{3 recursions}
& method & solver
\\\hline
$10^{5}$ & 1 & \textbf{0.704061} & 0.8647671 & 0.8682630 & 0.9244519
& 1.355239 & 1.848160\\
$10^{6}$ & 8 & \textbf{3.176656} & 3.301060 & 3.420162 & 3.458871
& 13.04333 & 4.715648\\
$10^{7}$ & 32 & 27.67224 & 26.04078 & 25.39645 & \textbf{25.07047}
& 125.2918 & 30.47072\\
$10^{8}$ & 32 & 271.9130 & 243.6584 & 242.6711 & \textbf{241.1610}
& 1236.283 & 283.8526\\\hline
\end{tabular}}
\end{table}

\begin{table}[!htbp]
\centering
\caption{Computational times (in [ms]) for the partition recursion methods on 
two NVIDIA GPU RTX A5000. The best time is given in bold font.}
\label{tab:recursive_MGPUs_pinned}
\resizebox{1\textwidth}{!}{%
\begin{tabular}{|c|c|c|c|c|c|r|}
\hline
\multicolumn{1}{|c|}{SLAE} & 
\multicolumn{1}{|c|}{number of} & 
\multicolumn{4}{|c|}{computational time for the partition method with} & 
\multicolumn{1}{|c|}{comparison with}\\ \cline{3-6}
\multicolumn{1}{|c|}{size $N$} & \multicolumn{1}{|c|}{CUDA streams} & 
\multicolumn{1}{|c|}{0 recursions} & \multicolumn{1}{|c|}{1 recursion} & 
\multicolumn{1}{|c|}{2 recursions} & \multicolumn{1}{|c|}{3 recursions} & 
\multicolumn{1}{|c|}{the time on 1 GPU}\\\hline
$10^{5}$ & 1 & \textbf{9.454705} & 9.925018 & 10.14414 & 10.17158 & worse than on 1 GPU \\
$10^{6}$ & 8 & \textbf{11.01846} & 11.69003 & 11.88673 & 12.25726 & worse than on 1 GPU \\
$10^{7}$ & 32 & 19.57763 & \textbf{18.94172} & 19.30156 & 19.89032 & better than on 1 GPU\\
$10^{8}$ & 32 & 118.6979 & 101.9684 & 101.5078 & \textbf{101.4709} & better than on 1 GPU\\
\hline
\end{tabular}}
\end{table}

\section{Conclusions}
This paper presents a multi-GPU version and a multi-GPU recursive version of the 
parallel partition method for systems of linear algebraic equations with 
tridiagonal coefficient matrices and their implementations in CUDA. Computational 
experiments on one and two GPU cards, for different system sizes and with 
different number of recursive steps are conducted.

For SLAE sizes $10^{5}$ and $10^{6}$ the best computational time comes on a 
single GPU, no recursions. 
Most probably the reason is that trying to use more than one GPU under-utilizes 
the resources, and the overhead is bigger than the gain. 
In particular, calling the CUDA API functions for setting the device, for 
synchronizing the streams, and the kernel launching across multiple GPUs carries an 
overhead. For big enough problems this overhead is amortized as it is negligible 
in comparison with the computational time. However, for relatively small 
problems this overhead might take longer than the computations, and thus not 
being able to be amortized. Additionally, if we try to split relatively small 
workloads, which do not fully saturate even one GPU, across multiple GPUs by 
cutting this workload into several chunks, instead of gaining performance, we 
end up running several even more under-utilized GPUs that spend most of their 
time waiting for warps to be scheduled, completely negating the benefit of having 
the compute power multiplied. Additional degradation of the performance in 
comparison with one GPU is caused by the modification of the for-loop workflow 
to stream all four SLAE arrays asynchronously. 
Finally, we can conclude that the result from the analysis of the optimum number 
of recursions which was made in~\cite{Veneva_2026} persists.

For SLAE size $10^{7}$ the best computational time comes on two GPUs, 
1 recursion. 
Taking into account the results for the optimum number of recursive steps 
from~\cite{Veneva_2026}, one would expect that 
the smallest computational time for this SLAE size should come when using 2 
recursive steps, because this is the optimum number of recursive steps for 
SLAE size $\approx5\times10^{6}$ which is the size that is computed per GPU. 
However, this is on the edge between 1 and 2 recursive steps. Hence, the result 
is not exceptional, and the result from the analysis of the optimum number of 
recursions persists. The achieved speedup in comparison with a single GPU is 
1.32 times. 
For SLAE size $10^{8}$ the best computational time comes on two GPU, 
3 recursions. Here, once 
again the result from the analysis of the optimum number of recursions persists. 
The achieved speedup in comparison with a single GPU is 2.38 times. This big 
speedup can be explained by the fact that for big enough workloads that are 
capable of fully saturating a single GPU, distributing the workload across two 
GPUs effectively halves the data footprint per card. Additional speedup is 
achieved by modifying the for-loop workflow to stream all four SLAE arrays 
asynchronously via the independent hardware copy engines. 
For both 
SLAE sizes $10^{7}$ and $10^{8}$ the worst computational time on two GPUs is 
better than the best time on a single GPU. For SLAE size $10^{7}$ the multi-GPU  
partition method demonstrates 69\% efficiency, while for SLAE size $10^{8}$ -- 
119\% efficiency when moving from one to two devices. Hence, these SLAE sizes 
are big enough to utilize both the GPUs. 

As of June 2026 (version 13.x), the cuSPARSE library~\cite{CUSPARSE} does 
not auto-parallelize across multiple GPUs. Therefore, we do not make comparison 
with its performance.

\section*{ACKNOWLEDGEMENTS}
I would like to thank Dr.~Toshiyuki Imamura (RIKEN) and Dr.~Alexander Ayriyan 
(JINR) for their precious comments.
\IEEEpeerreviewmaketitle

\end{document}